\documentclass[aps,pra,twocolumn,showpacs,superscriptaddress]{revtex4-1}

\usepackage{amsfonts,mathrsfs,amsmath,amsthm,amssymb}
\usepackage{epsfig,upgreek}
\usepackage{bm}
\usepackage{color}
\usepackage[dvipsnames]{xcolor}
\usepackage{adjustbox}
\usepackage{tabularx,bbding}
\usepackage{tabularx}

\usepackage{epsfig}
\usepackage{bm}
\usepackage{adjustbox}
\usepackage{tabularx}
\usepackage{tablefootnote}
\usepackage{booktabs}
\usepackage{multirow}
\usepackage{xfrac}
\usepackage{color}
\usepackage{graphicx}
\usepackage{tikz}
\usepackage{lipsum}
\usepackage{listings}
\usepackage{tabularx,bbding}

\pdfoutput=1
\usepackage[utf8]{inputenc}
\usepackage[english]{babel}
\usepackage[T1]{fontenc}
\usepackage{amsmath,ulem, amsfonts}
\usepackage{hyperref}
\usepackage{mathrsfs}%

\usepackage{tikz}
\usepackage{lipsum}

\usepackage{pifont}

\usepackage{enumitem}
\usepackage{graphicx}
\usepackage{xcolor}
\usepackage{braket}
\usepackage[ruled,vlined]{algorithm2e}

\usepackage{color, colortbl}
\definecolor{LightCyan}{rgb}{0.88,1,1}
\definecolor{Gray}{gray}{0.95}

\usepackage[colorinlistoftodos]{todonotes}
\definecolor{codegreen}{rgb}{0,0.6,0}
\definecolor{codegray}{rgb}{0.5,0.5,0.5}
\definecolor{codepurple}{rgb}{0.58,0,0.82}
\definecolor{backcolour}{rgb}{0.95,0.95,0.92}

\lstdefinestyle{mystyle}{
    commentstyle=\color{codegreen},
    keywordstyle=\color{magenta},
    numberstyle=\tiny\color{codegray},
    stringstyle=\color{codepurple},
    basicstyle=\ttfamily\footnotesize,
    breakatwhitespace=false,         
    breaklines=true,                 
    captionpos=b,   
    frame=lines,
    keepspaces=true,                 
    numbers=left,                    
    numbersep=5pt,                  
    showspaces=false,                
    showstringspaces=false,
    showtabs=false,                  
    tabsize=2,
    xleftmargin=1.8em
}

\makeatletter
\newsavebox{\@brx}
\newcommand{\llangle}[1][]{\savebox{\@brx}{\(\m@th{#1\langle}\)}%
 \mathopen{\copy\@brx\kern-0.5\wd\@brx\usebox{\@brx}}}
\newcommand{\rrangle}[1][]{\savebox{\@brx}{\(\m@th{#1\rangle}\)}%
 \mathclose{\copy\@brx\kern-0.5\wd\@brx\usebox{\@brx}}}
\makeatother

\begin{document}  
\title {\bf 
Multi-target quantum compilation algorithm
}

\author{Vu Tuan Hai}
\thanks{Electronic address: vu.tuan\_hai.vr7@naist.ac.jp}
\affiliation{Nara Institute of Science and Technology, 
Ikoma 630-0192, Nara, Japan}

\author{Nguyen Tan Viet}
\affiliation{FPT University, Hanoi, Vietnam}

\author{Jesus Urbaneja}
\affiliation{Department of Mechanical and Aerospace Engineering, Tohoku University, Sendai 980-0845, Japan}

\author{Nguyen Vu Linh}
\affiliation{University of Science,  Vietnam National University, 
Ho Chi Minh City 70000, Vietnam}
\affiliation{Vietnam National University, 
Ho Chi Minh City 70000, Vietnam}

\author{Lan Nguyen Tran}
\thanks{Electronic address:  tnlan@hcmus.edu.vn}
\affiliation{University of Science,  Vietnam National University, 
Ho Chi Minh City 70000, Vietnam}
\affiliation{Vietnam National University, 
Ho Chi Minh City 70000, Vietnam}

\author{ Le Bin Ho}
\thanks{Electronic address: binho@fris.tohoku.ac.jp}
\affiliation{Frontier Research Institute for Interdisciplinary Sciences, 
Tohoku University, Sendai 980-8578, Japan}
\affiliation{Department of Applied Physics, Graduate School of Engineering, Tohoku University, Sendai 980-8579, Japan}

\date{\today}

\begin{abstract}
Quantum compilation is the process of converting a target unitary operation into a trainable unitary represented by a quantum circuit. It has a wide range of applications, including gate optimization, quantum-assisted compiling, quantum state preparation, and quantum dynamic simulation. Traditional quantum compilation usually optimizes circuits for a single target. However, many quantum systems require simultaneous optimization of multiple targets, such as thermal state preparation, time-dependent dynamic simulation, and others. To address this, we develop a multi-target quantum compilation algorithm to improve the performance and flexibility of simulating multiple quantum systems. Our benchmarks and case studies demonstrate the effectiveness of the algorithm, highlighting the importance of multi-target optimization in advancing quantum computing. This work lays the groundwork for further development and evaluation of multi-target quantum compilation algorithms.
\end{abstract}
%
%
\maketitle

\section{Introduction}
\label{sec1}
Variational quantum algorithms (VQAs) 
like quantum approximate optimization algorithm (QAOA), 
variational quantum eigensolver (VQE),
quantum neural networks (QNN),
and quantum compilation (QC) 
are promising for solving practical tasks 
on noisy intermediate scale quantum (NISQ) 
devices beyond classical computers \cite{Cerezo2021}. 
VQAs have proven to be versatile in various applications, including
quantum state preparation \cite{
Kuzmin2020variationalquantum,
Sagastizabal2021,
PhysRevLett.129.230504,
castro2024variational,
hai2023variational,HAI2024101726}, 
quantum dynamic simulation 
\cite{HAI2024101726,PRXQuantum.2.030307,
Luo2024,PhysRevResearch.6.023130}, 
and quantum metrology \cite{Koczor_2020,9605341,
Meyer2021,Le2023,Cimini2024}. 
Recent developments underscore their ability to tackle complex quantum systems effectively.

In quantum state preparation, significant progress has led to various effective methods for generating target states with high fidelity \cite{Kuzmin2020variationalquantum,
Sagastizabal2021,
PhysRevLett.129.230504,
castro2024variational,
hai2023variational,HAI2024101726}. 
For instance, Zhang et al. demonstrated that any quantum state can be prepared with a linear-depth circuit using many ancillary qubits, leading to exponential speedups for tasks like Hamiltonian simulation and solving linear systems \cite{PhysRevLett.129.230504}. Additionally, VQAs have been extensively studied for simulating quantum dynamics, employing adaptive variational principles to optimize the time evolution of quantum states \cite{PRXQuantum.2.030307}. This approach is particularly important for simulating open quantum systems, where the interaction between system dynamics and environmental factors plays a crucial role \cite{PRXQuantum.5.020332}. Furthermore, VQAs have demonstrated their capability to achieve quantum-enhanced precision in multiparameter quantum metrology, even in the presence of noise \cite{Le2023}.

Quantum compilation, on the other hand, 
has gained significant interest
due to its capacity to optimize 
quantum circuits through 
a training process that transforms 
target unitaries into trainable unitaries
\cite{heya2018variational,Khatri2019quantumassisted}. 
This approach has been applied to various tasks, 
including gates optimization \cite{heya2018variational}, 
quantum-assisted compiling \cite{Khatri2019quantumassisted}, 
continuous-variable quantum learning \cite{PRXQuantum.2.040327}, 
quantum state tomography \cite{hai2023universal}, 
and quantum objects simulation \cite{HAI2024101726}.
For instance, a quantum object such as a quantum state 
can be prepared and its evolution can be simulated in a quantum circuit by using QC \cite{HAI2024101726}.

The performance of QCs relies on the number of qubits and the circuit depth. 
The choice of ansatzes (trainable quantum circuits) is also crucial 
and must be carefully selected. Some entangled topologies have shown 
promise in solving quantum state preparation problems 
but require many resources due to their numerous layers 
and parameters \cite{hai2023variational}. 

Traditional QCs 
focus on compiling a single target
\cite{heya2018variational,Khatri2019quantumassisted,PRXQuantum.2.040327,hai2023universal,HAI2024101726} as mentioned above.
However, many quantum systems require optimizing multiple factors at the same time, because independent optimization of each target can lead to suboptimal results in complex scenarios. For instance, simulating time-dependent quantum systems \cite{PhysRevA.99.042314,PRXQuantum.2.030307,Berry2024quantumalgorithm,Mizuta2023optimalhamiltonian} often requires balancing accuracy across various time intervals, while thermal-dependent systems \cite{Sagastizabal2021,PhysRevLett.123.220502,Zhu_2020,PhysRevApplied.16.054035} demand simultaneous optimization of thermal fluctuations and quantum coherence to maintain system stability. Similarly, in multimode tomography \cite{PhysRevA.90.053818,He2024}, optimizing multiple modes concurrently enhances the overall precision of quantum state estimation. These examples demonstrate that multi-target optimization is crucial in achieving optimal performance in quantum applications. 

This work presents a multi-target quantum compilation algorithm 
to address these demains by providing a unified framework for optimizing multiple objectives simultaneously.
Therein, we compile multiple unitaries into a single 
trainable unitary through an optimization process, 
improving the performance of QC and applicability 
in practical scenarios. 
We explain the theoretical
foundations of our algorithm, its structure, 
and using techniques like the genetic algorithm (GA) 
to enhance the multi-target QC process. 
We show the effectiveness of our algorithm 
through benchmarking and case studies, 
including the preparation of thermal states and the simulation of time-dependent Hamiltonians
in comparison with other methods including  
Trotterlization \cite{Ikeda2023minimum,ikeda2023trotter24}
and adaptive variational quantum dynamics simulations (AVQDS) \cite{PRXQuantum.2.030307}.
We also provide a demonstration for the variational quantum eigensolver.
The results highlight the importance of 
multi-target compilation in advancing 
quantum computing technologies and the potential 
for further innovation in this field.

The novelty of this work includes:
\begin{itemize}
    \item  It presents a quantum compilation algorithm for the first time that is adaptable for simulating multiple unitaries.
    \item It introduces an enhanced quantum architecture search that utilizes a genetic algorithm in combination with a standard variational quantum algorithm.
\end{itemize}

The paper is structured as follows. In Section~\ref{sec2}, we present the multi-target quantum compilation with detail of  its definitions, algorithms, and numerical benchmarking. Section~\ref{sec3} explores various applications, including thermal state preparation, time-dependent quantum dynamic simulations, and variational quantum eigensolvers. Finally, we conclude the paper in Section~\ref{sec4}.
Additional material can be found in \ref{appA} and \ref{appB}.

\section{Multi-target quantum compilation}
\label{sec2}
Quantum compilation (QC) 
\cite{Khatri2019quantumassisted, hai2023variational} 
can be regarded as
 a training process 
that transforms information 
from a target unitary $U$ 
to a trainable unitary 
$V(\bm{\theta})$, 
where $\bm{\theta}$ 
are trainable parameters.
Hereafter, we introduce a compilation 
process to transform a set of $n$ targets
into one trainable unitary (one single quantum circuit with $n$ different $\bm\theta$s), therefore, 
it is called multi-target quantum compilation.

\subsection{Definition}
Let $\bm{\mathscr{H}}$ be a Hilbert space, and  
$\mathcal{U} = \big\{ U_j ; 1 \le j \le n\big\}$ 
be $n$-target unitaries on $\mathscr{H}$,
there exists a trainable unitary $V$
that satisfies
\begin{align}\label{eq:UVd}
    U_jV^\dagger(\bm\theta_j) 
    = e^{-i\phi}\mathbb{I}, \; 
    \forall 1 \le j \le n,
\end{align}
where $\bm\theta_j \in \big\{\theta_j^{(1)}, 
\theta_j^{(2)}, ..., \theta_j^{(m)}\big\}$
is a set of $m$ trainable parameters
and $\phi$ is a global phase. For each $U_j$, 
we need to maximize the kernel 
$\mathcal{K}_{U_j,V}(\bm\theta_j)
=\frac{1}{u^2}\left|{\rm Tr}[U_j 
V^{\dagger}(\bm{\theta}_j)]
\right|^2$ 
where $u$ is the dimension of $U_j$.
For maximizing all 
$\mathcal{K}_{U_j,V}(\bm\theta_j), 
\forall U_j \in \mathbf{\mathcal{U}}$, 
we define a cost function $\mathcal{L}$,
which is the average infidelity 
\begin{align}\label{eq:f_qc}
    \mathcal{L}(\bm \Theta, V) 
    =1-\frac{1}{n} \sum_{j=1}^{n}
    \mathcal{K}_{U_j,V}(\bm\theta_j),
\end{align}
where $\bm\Theta = (\bm\theta_1, \cdots, 
\bm\theta_n)$ are trainable parameters, and 
$0 \leq \mathcal{L}(\bm\Theta,V) \leq 1$.
The optimization process is to find
$
    \bm\Theta^*, V^* = 
    \operatorname*{argmin}_{\{\bm\Theta, V\}}
    \mathcal{L}(\bm\Theta, V).
$
Here, we emphasize that not only the trainable parameters are optimized
but also the structure of ansatz $V$ is also optimized.

\subsection{Quantum compilation algorithm}
The training process in QC 
is a variational quantum algorithm (VQA).
Given a target unitary $U$, 
it employs a gradient-based optimizer 
to update an ansatz $V(\bm\theta)$
and find the optimal $\bm\theta^*$. 
The ansatz $V$ is usually a multi-layer structure
\begin{align}\label{eq:v_decompose}
    V(\bm \theta) = \prod_{k=1}^L 
    G_l(\bm \theta_l) \in \text{SU}(2^N),
\end{align}
where $G_k(\bm \theta_k)$ is a sequence of parameterized 
single- and multi-qubit quantum gates, $L$ is the number of 
layers, $N$ the number of qubits.
The design of $V(\bm \theta)$ is not unique, 
and recent efforts to optimize 
it, including architecture search 
\cite{Du2022,Zhang_2021,PhysRevResearch.2.023074,
https://doi.org/10.1002/qute.202100134}, 
adaptive variational quantum algorithm \cite{Grimsley2019,PRXQuantum.2.030307},
and genetic-based approach (GA)
\cite{ 934383,8862255,katoch2021review,Tandeitnik_2024}.

\begin{figure*}[t]
\includegraphics[width=\textwidth]{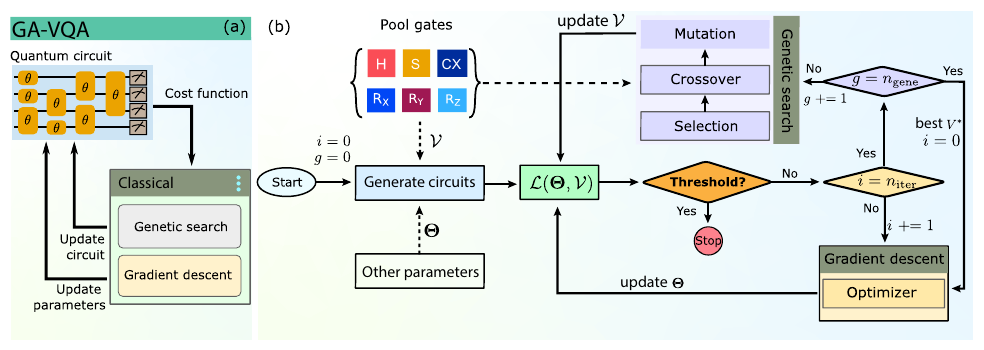}
\caption{
\textbf{Integrated Genetic Algorithm with Variational Quantum Algorithm (GA-VQA) for multi-target quantum compilation optimization.}
(a) The GA-VQA method uses a parameterized quantum circuit and a classical computer to evaluate and update both the circuit structure and its parameters.
(b) The process is as follows:
(i) Generate a set of circuits $\mathcal{V}$ and parameters $\bm\Theta$.
(ii) Evaluate the cost function $\mathcal{L}(\bm\Theta, V_l)$ for each circuit $l \in [1, n_\mathcal{V}]$.
(iii) If the threshold is not met, use VQA followed by GA to find the best circuit, checking the threshold after each iteration.
(iv) If the threshold is still not reached, pass the best circuit $V^*$ back to VQA and repeat the process until the threshold is met.
Here, $i$ and $g$ represent iteration indexes running up to $n_{\rm iter}$ and $n_{\rm gene}$, respectively.
}
\label{fig:1}
\end{figure*}

In this work, we combine GA with VQA to optimize multi-target quantum compilation. This approach optimizes both the parameters and the structure of quantum circuits (ansatzes), ensuring efficient compilation, practical computational capability, and low-depth circuits.
The scheme is shown in Figure \ref{fig:1}(a), 
and the flow chart is given in Figure \ref{fig:1}(b).
We first generate quantum circuits with random parameters and evaluates their cost functions. If the cost functions do not meet a threshold, we use VQA to optimize the parameters and re-evaluate. After VQA finishes, the best circuits are passed to the GA, which updates their structure through selection, crossover, and mutation for the next generation. Once GA completes, the best circuit is fed back into the VQA, which runs until the threshold is met or the cost function stabilizes.

Concretely, denote the set of generated quantum circuits as $\mathbf{\mathcal{V}} = \big\{V_l; 1 
\le l \le n_\mathbf{\mathcal{V}}\big\}$,
where $n_\mathbf{\mathcal{V}}$ is the number of circuits. Each circuit $V_l \in \mathbf{\mathcal{V}}$ is created from a pool of quantum gates.
For each circuit $V_l$, we evaluate the average infidelity via
$\mathcal{L}(\bm\Theta, V_l)\ \forall 1 \le l \le n_\mathbf{\mathcal{V}}. $
If $\mathcal{L}(\bm\Theta, V_l) 
\le \rm threshold$, we designate $V_l$ 
to the best circuit, $V^* \leftarrow V_l$, 
and $\bm\Theta$ will be the optimal parameters, 
$\bm\Theta^* \leftarrow \bm\Theta$.
If not, we run the VQA scheme with 
gradient descent to update $\bm\Theta$ 
for a certain number of iterations $n_{\rm iter}$.
Then, we use GA with selection, crossover, 
and mutation to create the next generation 
of $\mathcal{V}$. See \ref{appA} for a detailed 
evolution of the GA.
The GA scheme repeats until the threshold 
is met or the number of 
generations $n_{\rm gene}$ is reached.
Finally, if $n_{\rm gene} $ is reached 
without meeting the threshold, 
we pass the best circuit $V^*$ 
to the VQA scheme again, 
and optimize $\bm\Theta$ 
until the threshold is met. 

In this framework, we run the VQA process twice. 
The first run aims to create a suitable
$\mathcal{L}(\bm\Theta, V_l)$ 
for evaluating the GA evolution process.
We can choose a small $n_{\rm iter}$ 
in this step to reduce computational cost. The second run occurs after the GA and aims to optimize $\bm\Theta$ 
to ensure that 
$\mathcal{L}(\bm\Theta^*, V^*)$ 
is minimized within the best circuit $V^*$.

Our scheme differs from the one-stage 
\cite{Du2022} and two-stage architecture search \cite{PhysRevResearch.2.023074,Ostaszewski2021structure}. 
In those methods, all circuits are generated, their cost functions are evaluated as the low-depth and high number of gates as much as possible (gate density \cite{10.1145/3550488} close to unity), and they are sorted from best to worst to select the top one. In our approach, each circuit $V_l\in\mathcal{V}$ 
is evaluated and compared with the best circuit (initially set as $V_1$).
We thus do not require space to store all circuits.
Although genetic algorithms have been suggested for quantum state tomography \cite{Creevey2023}, we believe they are also promising for different types of VQAs, including our proposed method for quantum compilation.

So far, we have observed that Ashhab et al. found that if a quantum circuit can perfectly implement one arbitrary target using random search, it can also implement any other targets with the same gate configuration. It only requires recalculating the single-qubit rotation parameters for each new target \cite{PhysRevA.106.022426,PhysRevA.109.052605}. This finding is similar to our discovery here using quantum compilation.

\subsection{Numerical illustration}
For numerical evaluation,
we follow Ref.~\cite{Caro2023} and divide a random set 
$\mathcal{U}$ into a training set 
$\mathcal{U}_{\text{train}}=\{U_1,U_2, ..., U_{\rm train} \}$ 
and a testing set $\mathcal{U}_{\text{test}}=\{U_{{\rm train + 1}},
U_{\rm train + 2}, ..., U_{n} \}$. 
We then train the model on $\mathcal{U}_{\text{train}}$ 
using the cost function $\mathcal{L}(\bm \Theta_{\rm train}, V_l)$
as in Eq.~\ref{eq:f_qc}.
After training, we get the best circuit $V^*$ 
and evaluate the expected risk $\mathcal{R}$
over $\mathcal{U}_{\text{test}}$
\begin{align}\label{eq:risk_qc}
\notag \mathcal{R}(\bm\Theta_{\rm test})&=\frac{1}{4} \mathbb{E}_{U_j\;\in\;\mathcal{U}_{\text {test}}}
\Big\|U_j|\bm 0\rangle\langle \bm 0|U_j^{\dagger} \\
&\hspace{2cm}-[V^*]^{\dagger}(\bm\theta_j) |\bm 0\rangle\langle \bm 0 | V^*(\bm\theta_j)\Big\|_1^2,
\end{align} 
where 
$|\bm 0\rangle 
\equiv |0\rangle^{\otimes N}$,
with $N$ is the number of qubits.

\begin{figure*}[t]
\includegraphics[width=16.6cm]{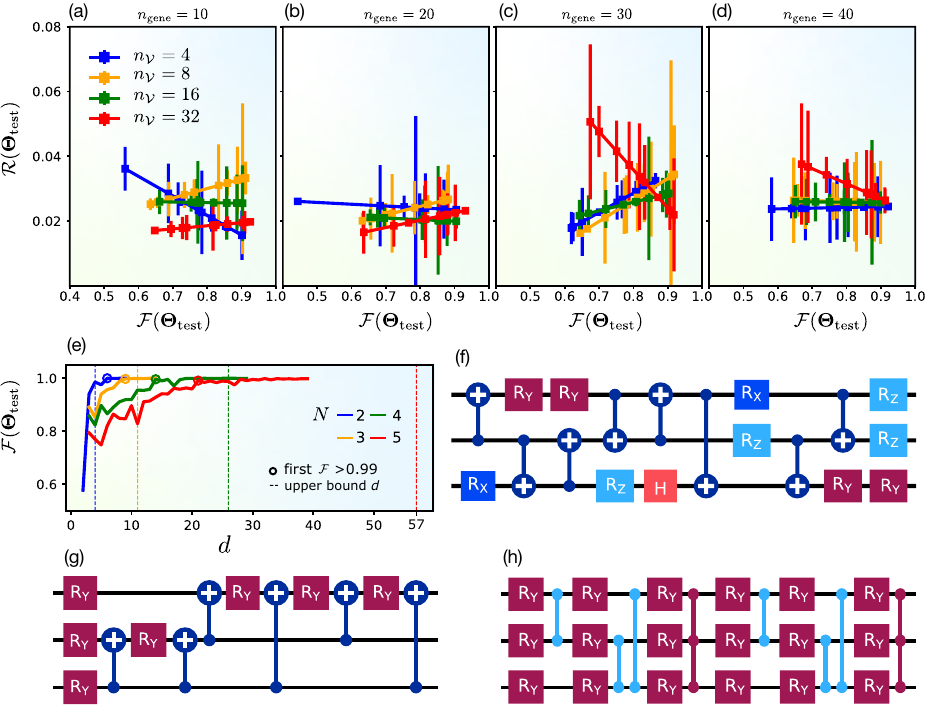}
\caption{
\textbf{Numerical benchmarking for 
a set of Haar random $\mathcal{U}$.}
(a-d) Correlation between risk and 
fidelity for different number of 
generations $n_{\rm gene} $ and circuits per generation $n_{\mathcal{V}}$.
(e) Plot of fidelity versus depth $d$ 
for different number of qubits $N$ ranging from 2 to 5.
(f) The best quantum circuit generated by GA-VQA for $N = 3$ with $d = 9$.
(g) The quantum circuit generated by Qiskit with $d = 11$.
(h) $|g_2 g_N\rangle$ (2 layers) 
ansatz before transforming into single and two-qubit gates.}
\label{fig:2}
\end{figure*}

Figure ~\ref{fig:2}(a-d) shows the correlation 
between risk $\mathcal{R}$ and fidelity 
$\mathcal{F} = 1-\mathcal{L}$ 
for several $n_{\rm gene} $ and $n_{\mathcal{V}}$ 
with $N = 3$. The GA-VQA method identifies a low-risk, high-fidelity ansatz for multi-target Haar random unitaries. Additionally, the factors $n_{\rm gene} $ and 
$n_{\mathcal{V}}$ have minimal impact on the results, allowing GA-VQA to run efficiently with minimal values, thus saving computational resources.
The variation in some cases, such as the yellow lines in Figure ~\ref{fig:2}(a-c), is a result of fluctuations in the optimization process for specific parameter settings. These fluctuations can arise from the randomized nature of the initial population in the genetic algorithm or the stochastic behavior of the variational quantum algorithm (VQA). Furthermore, the performance of different circuit architectures can vary significantly, leading to occasional deviations in the optimization trajectory. 

Figure ~\ref{fig:2}(e) shows fidelity versus depth $d$ for 2-5 qubits. As expected, higher $N$ requires larger $d$ to achieve high fidelity. 
In the figure, circles mark the depth where $\mathcal{F} \ge 0.99$, with upper bounds generated by the standard Qiskit method.
Figure ~\ref{fig:2}(f) shows the best circuit generated by our GA-VQA compared to the Qiskit-generated circuit in Figure ~\ref{fig:2}(g) and the $|g_2g_N\rangle$ ansatz\cite{hai2023variational} in Figure ~\ref{fig:2}(h) for $N = 3$. The GA-VQA circuit has the shortest depth (9), compared to 11 and 38 for the others. Table~\ref{tab:compare-qc} shows detailed comparisons for 2 to 5 qubits, demonstrating that the GA-VQA circuit provides high fidelity with the smallest depth.

\begin{table}[ht]
\caption{Comparison between three methods: 
GA-VQA, default Qiskit preparation, 
and $|g_2 g_N\rangle$ (2 layers). 
All circuits are transpiled 
with the same gate set 
$\{{\rm H, S, CX, R}_i(\theta)\}$ with $i\in\{x,y,z\}$.}
\label{tab:compare-qc}
\begin{adjustbox}{width=\columnwidth,center}
\begin{tabular}{|l|lll|lll|lll|lll|}
\hline
\multicolumn{1}{|c|}{$N$} & \multicolumn{3}{l|}{$\mathcal{F}$} & \multicolumn{3}{l|}{Depth ($d$)} & \multicolumn{3}{l|}{\#gates (\#1-qubit gate + \#2-qubit gate)} & \multicolumn{3}{l|}{\#parameters} \\ \hline
 & \multicolumn{1}{l|}{GA-VQA} & \multicolumn{1}{l|}{Qiskit} & $|g_2 g_N\rangle$ & \multicolumn{1}{l|}{GA-VQA} & \multicolumn{1}{l|}{Qiskit} & $|g_2 g_N\rangle$ & \multicolumn{1}{l|}{GA-VQA} & \multicolumn{1}{l|}{Qiskit} & $|g_2 g_N\rangle$ & \multicolumn{1}{l|}{GA-VQA$^*$} & \multicolumn{1}{l|}{Qiskit} & $|g_2 g_N\rangle$ \\ \hline
2 & \multicolumn{1}{l|}{0.92} & \multicolumn{1}{l|}{0.99} & 0.99 & \multicolumn{1}{l|}{3} & \multicolumn{1}{l|}{4} & 24 & \multicolumn{1}{l|}{4+1} & \multicolumn{1}{l|}{3+2} & 24+6 & \multicolumn{1}{l|}{5} & \multicolumn{1}{l|}{3} & 12 \\ \hline
3 & \multicolumn{1}{l|}{0.99} & \multicolumn{1}{l|}{0.99} & 0.99 & \multicolumn{1}{l|}{9} & \multicolumn{1}{l|}{11} & 38 & \multicolumn{1}{l|}{11+8} & \multicolumn{1}{l|}{7+6} & 40+18 & \multicolumn{1}{l|}{17} & \multicolumn{1}{l|}{7} & 18 \\ \hline
4 & \multicolumn{1}{l|}{0.99} & \multicolumn{1}{l|}{0.99} & 0.99 & \multicolumn{1}{l|}{15} & \multicolumn{1}{l|}{26} & 97 & \multicolumn{1}{l|}{32+14} & \multicolumn{1}{l|}{15+14} & 104+52 & \multicolumn{1}{l|}{40} & \multicolumn{1}{l|}{15} & 24 \\ \hline
5 & \multicolumn{1}{l|}{0.99} & \multicolumn{1}{l|}{0.99} & 0.97 & \multicolumn{1}{l|}{27} & \multicolumn{1}{l|}{57} & 166 & \multicolumn{1}{l|}{63+36} & \multicolumn{1}{l|}{31+30} & 158+98 & \multicolumn{1}{l|}{73} & \multicolumn{1}{l|}{31} & 30 \\ \hline
\hline
\multicolumn{13}{l}
{$^*$The number of parameters in GA-VQA 
differ for each run depending on the set of gates in the best circuit.}\\
\end{tabular}
\end{adjustbox}
\end{table}

\subsection{The complexity of GA-VQA}
We discuss here the complexity of GA-VQA.
Time complexity describes how the runtime of an algorithm changes as the input size increases, usually expressed using Big-O notation. It gives an estimate of the maximum runtime as the input grows. Common complexities include $\mathcal{O}(1)$ for constant time, $\mathcal{O}(z)$ for linear time, $\mathcal{O}(z^2)$for quadratic time, and $\mathcal{O}(\log z)$ for logarithmic time. 
Here, $z$ is the input size.
The time complexity of GA-VQA is defined as 
$\mathcal{O}(n \times n_{\rm gene} \times n_{\mathcal{V}} \times \mathcal{F})$ 
which depends on 
the number of targets $n$, 
the number of generations 
$n_{\rm gene}$, 
the number of circuits in each generation 
$n_{\mathcal{V}}$, and 
the 
fidelity $\mathcal{F}$. 
When the fidelity of circuits in 
one generation can be evaluated concurrent (parallel mode), 
the actual time complexity is 
$\mathcal{O}(n_{\rm gene} 
\times n_{\mathcal{V}}\times \mathcal{F})$, 
assuming that number of processing cores is larger than $n$.

The space complexity of an algorithm refers to the amount of memory or storage it requires relative to the size of the input. 
The space complexity in our case is $O(n \times n_{\mathcal{V}}  \times \mathcal{F})$.

Since \( n_{\rm gene}, n_{\mathcal{V}}\) are hyper-parameters, the time and space complexities heavily depend on the cost function. For instance, for GA-VQA applied to multi-target compilation, the time complexity is approximately $\mathcal{O}(n_{\rm gene} \times n_{\mathcal{V}} \times n_{\text{iter}} \times 2^N \times d
)$, where \( d \) is the circuit depth and \( n_{\text{iter}} \) is the number of optimization iterations. The space complexity for a state vector simulator is at least \( \mathcal{O}(n \times n_{\mathcal{V}} \times 2^N) \). For example, with 8 GB of RAM, GA-VQA can support up to 10 qubits with \( n = 5 \) and \( n_{\mathcal{V}} = 4 \), which results in approximately 6 GB used, calculated as follows: \( 5 \times 4 \) (for concurrent processes), \( 64 \times 2 \times (2^{10} + 2^{20}) \) (for a double-precision, 10-qubit quantum state-operator), multiplied by 2 for each buffer (state and circuit).

Currently, GA-VQA can manage unitaries that act on around 8-10 qubits with moderate circuit depths of approximately 40. More details can be found in \ref{appB}. However, this estimate may change with advancements in hardware capabilities, such as improved qubit coherence times and error rates.
These factors are also the bottleneck for scaling the system size.

\section{Applications}
\label{sec3}
\subsection{Thermal state preparation (TSP)}
Quantum state preparation using quantum algorithms 
is well-studied for pure states 
\cite{Kuzmin2020variationalquantum,castro2024variational,hai2023variational}. 
However, preparing mixed states requires purification first, followed by the preparation of the pure state \cite{ezzell2023quantum}. 
The conventional purification method
needs $2N$ qubits to prepare a mixed state of $N$ qubits.
Here, we propose a ``dense-purification'' method that needs only $N$ qubits.

A quantum state $\rho$ in a $u$-dimensional 
Hilbert space $\mathscr{H}_A$ can be represented 
using its eigenvalues and eigenstates as 
$\rho = \sum_{j=1}^u p_j |j\rangle\langle j|$.
In conventional purification, we create a pure state
$|\psi\rangle$ in a larger Hilbert space 
$\mathscr{H}_A\otimes \mathscr{H}_B$,
where $\mathscr{H}_B$ is another $u$-dimensional Hilbert space
with an orthonormal basis $\{|j\rangle_B\}$.
The pure state is $|\psi\rangle = \sum_j \sqrt{p_j} 
|j\rangle_A|j\rangle_B$, purifying $\rho$ 
such that $\rho = {\rm Tr}_B[|\psi\rangle\langle\psi|]$.
In our dense-purification method, we define a pure state 
$|\psi\rangle = \sum_j \sqrt{p_j}|j\rangle$, 
directly representing the dense-purified state of $\rho$. 
It can be shown that $\rho = {\rm diag}(|\psi\rangle\langle\psi|)$, 
allowing us to extract various properties of $\rho$ from $|\psi\rangle$.

We demonstrate the preparation of thermal equilibrium states, 
specifically Gibbs states at fixed temperatures. 
These states are crucial for various applications, 
such as quantum simulation \cite{9571985}, 
quantum machine learning \cite{PhysRevX.8.021050}, 
quantum condensed matter \cite{PRXQuantum.4.010305}, 
quantum field theory \cite{Lewin2021}, and cosmology 
\cite{ISRAEL1976107,Gao2017}. We focus on the Gibbs 
state of a transverse field Ising model (TFIM) 
\cite{10.21468/SciPostPhys.6.3.029}, which is 
useful for studying thermal phase transitions 
in condensed matter physics.

\begin{figure*}[t]
\includegraphics[width=\textwidth]{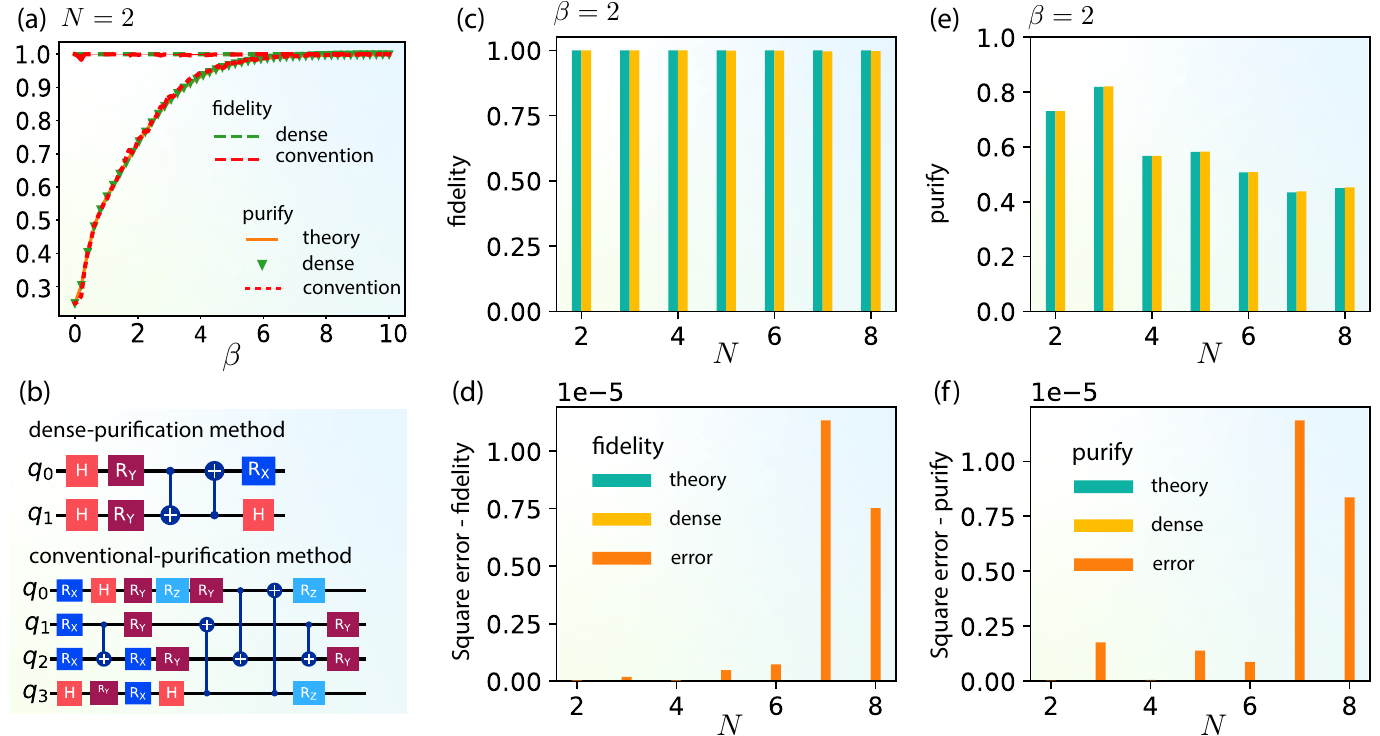}
\caption{
\textbf{Thermal states preparation.}
(a) Comparison of fidelity and purity versus $\beta$ at $N = 2$
for various methods, alongside theoretical predictions.
(b) Quantum circuits used in dense and conventional methods.
(c) Fidelity plotted against $N$ at $\beta = 2$.
(d) Square error of fidelity plotted against $N$ at $\beta = 2$.
(e) Purity plotted against $N$ at $\beta = 2$.
(f) Square error of purity plotted against $N$ at $\beta = 2$.
The legends for (c) and (e) are the same as (d) and (f).
}
\label{fig:3}
\end{figure*}

In the TFIM model on a ring of $N$ sites, 
the Hamiltonian takes the form
$H = \sum_{i=1}^{N} Z_{i}Z_{i+1}+\sum_{i=1}^{N} X_i$, 
where $X$ and $Z$ are Pauli matrices.
In this model, the Gibbs state is defined as
$\label{eq:Gibbs}
\rho(\beta) = e^{-\beta H} / Z(\beta)
$
where $\beta = k_BT$ is the inverse temperature 
and $Z(\beta) = {\rm Tr}[e^{-\beta H}]$ is the partition function.
At $\beta = 0$, the Gibbs state is maximally mixed, 
and it gradually becomes a pure state when $\beta \to \infty$.

Conventionally, to prepare this mixed state on a quantum computer, 
previous approaches used the conventional purification scheme as
\cite{Sagastizabal2021,PhysRevLett.123.220502,Zhu_2020,PhysRevApplied.16.054035}
\begin{equation}\label{eq:pure}
    |\psi(\beta)\rangle = \frac{1}{\sqrt{Z(\beta)}}
    \sum_{j}e^{-\beta E_j/2}|j\rangle_A|j\rangle_B,
\end{equation}
where $Z(\beta) = \sum_{j}e^{-\beta E_j}$.
Here, $E_j$ and $|j\rangle$ are 
the eigenvalues and eigenstates of $H$, 
i.e., $H|j\rangle = E_j|j\rangle$.
The purified state is known as
thermofield double (TFD) state \cite{Sagastizabal2021},
which related to the Gibbs state through  
$\rho(\beta) = {\rm Tr}_B
\big[|\psi(\beta)\rangle\langle\psi(\beta)|\big]$.

For the dense-purification approach, 
the dense-purified state is given by
\begin{equation}\label{eq:dense-pure}
    |\psi_{\rm dense}(\beta)\rangle = 
    \frac{1}{\sqrt{Z(\beta)}}
    \sum_{j}e^{-\beta E_j/2}|j\rangle_A.
\end{equation}

Previously, QAOA \cite{Premaratne_2020,PhysRevLett.123.220502,Zhu_2020,Sagastizabal2021}
and parametrized circuits~\cite{PhysRevApplied.16.054035} were used to prepare the quantum state in \eqref{eq:pure}. 
Here, we apply our GA-VQA to prepare the states in Eqs. (\ref{eq:pure}, \ref{eq:dense-pure}).
We evaluate the closeness of the prepared state to the target state using fidelity and purity metrics
\begin{equation}\label{eq:GibbsFidelity}
    \mathcal{F} = {\rm Tr}\Big[\sqrt{\sqrt{\rho(\beta)}\check{\rho}(\beta)\sqrt{\rho(\beta)}}\Big]^2; \ 
    \text{ and }\mathcal{P} = {\rm Tr}[\check{\rho}(\beta)],
\end{equation}
where $\check{\rho}(\beta) = {\rm Tr}
\big[|\psi(\beta)\rangle\langle\psi(\beta)|\big]$
for conventional purification and $\check{\rho}(\beta) = {\rm diag}
\big[|\psi_{\rm dense}(\beta)\rangle\langle\psi_{\rm dense}(\beta)|\big]$
for dense purification.
See \ref{appB} for details on preparing $|\psi(\beta)\rangle$
and $|\psi_{\rm dense}(\beta)\rangle$. 

In Figure ~\ref{fig:3}(a), we set $N = 2$ and examine $\mathcal{F}$
and $\mathcal{P}$ while comparing them in different methods.
We demonstrate that both methods align well with the theory, 
proving the efficacy of the preparation method for creating Gibbs states.
In Figure ~\ref{fig:3}(b), we show 
the corresponding quantum circuits, 
where the dense method requires fewer qubits 
and lower circuit depth compared to the conventional method.
Next, we focus on the dense method with $\beta = 2$. 
Figure \ref{fig:3}(c) shows the fidelity 
versus the number of qubits $N$ with 
the results closely approaching one (theory) 
for all $N$, and the square error is 
on the order of $10^{-5}$ as depicted 
in Figure ~\ref{fig:3}(d).
Similarly, we analyze the purity and 
its error in Figure ~\ref{fig:3}(e,f), 
which align well with the theory.

\subsection{Time-dependent quantum dynamic simulation (TD-QDS)}
In this section, we demonstrate 
for dynamic time-dependent simulations. 
We consider a one-dimensional spin-1/2 system with $N$ spins, 
initially prepared in a domain wall configuration
$|\psi_0\rangle = |\cdots \downarrow\downarrow\uparrow\uparrow
\cdots\rangle$.
The time-dependent Hamiltonian is given by
\begin{widetext}
\begin{align}\label{eq:XYZ_model}
    H(t) = -\frac{J}{2}\sum_{j=1}^{N-1}
    \left[\Big(1-\dfrac{t}{T}\Big) X_j X_{j+1} + 
    \Big(1+\dfrac{t}{T}\Big) Y_jY_{j+1} \right]+ 
    u \sum_{j=1}^{N-1} Z_jZ_{j+1} + 
    h \sum_{j=1}^{N} X_j,    
\end{align}
\end{widetext}
where $J$ and $u$ are coupling strengths,
$h$ is the external magnetic field,
and $X, Y, Z$ are Pauli matrices. 
The evolution is given by
\begin{align}\label{eq:UXYZ}
    U(t) = \mathcal{T}
    \exp\Big(-i\int_0^t ds H(s)\Big),
\end{align}
where $\mathcal{T}$ is the time-ordering operator, 
and $\hbar = 1$ is used throughout the paper. 
We consider the local magnetization 
as the dynamical quantity to be examined
\begin{align}\label{eq:M_theo}
M_j(t) =  
\langle\psi(t)|Z_j|\psi(t)\rangle,
\end{align}
where $|\psi(t)\rangle$ is 
the quantum state given at time $t$.

\begin{figure*}[t]
\centering
\includegraphics[width=\linewidth]{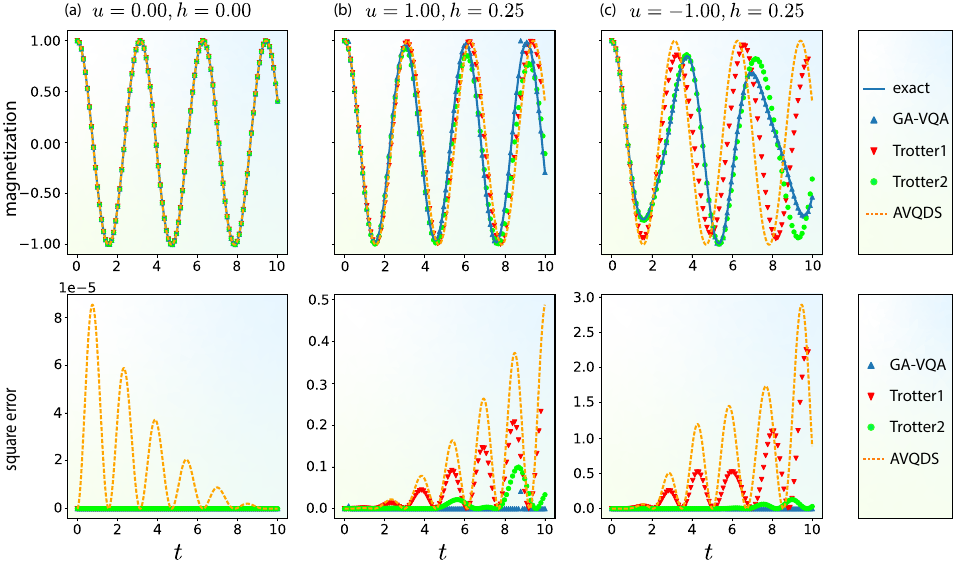}
\caption{
\textbf{Time-dependent quantum dynamic simulation}.
Plot of the local magnetization versus time (upper)
and its square error (lower) for several models:
(a) $u = 0.00, h = 0.00$,
(b) $u = 1.00, h = 0.25$,
(c) $u = -1.00, h = 0.25$.
Here we set $N =  2$, $J = 1$ and $T = 10$.
We compare various methods with the theory, including 
GA-VQA, Trotter1, Trotter2, and AVQDS.
}
\label{fig:4}
\end{figure*}

The local magnetization is shown 
in Figure ~\ref{fig:4}, 
comparing the exact result with various methods, including GA-VQA, Trotterlization \cite{Ikeda2023minimum,ikeda2023trotter24}
with first order (Trotter1) and second order (Trotter2), and adaptive variational quantum dynamics simulations (AVQDS) \cite{PRXQuantum.2.030307}. 
We set $N = 2, J = 1, T = 10$ 
and consider several models 
for $u$ and $h$ as shown in the figure.
For the theoretical computation, 
the local magnetization is given by 
Eq.~\eqref{eq:M_theo}, 
with the final state directly computed from 
$U(t)$ yielding $|\psi(t)\rangle = U(t)|\psi_0\rangle$.
For the other simulation methods, the local magnetization
is derived from the measured probabilities of the final circuit. See detailed calculation in \ref{appB}.

The GA-VQA results closely match the exact results over time for all models, demonstrating the method's reliability and versatility in capturing quantum dynamics.
Trotter2 also achieves high accuracy, aligning well with exact results in cases (a) and (b), but shows deviations in case (c).
Trotter1 and AVQDS match well with theoretical predictions only in case (a) and deviate as time 
$t$ increases in cases (b) and (c). Deviations in Trotter1 are common, as noted in previous studies \cite{Ikeda2023minimum,zhao2023adaptive,ikeda2023trotter24}, while discrepancies in AVQDS results stem from simplifications in our calculations, as discussed in \ref{appB}.

\subsection{Demo application for Variational Quantum Eigensolver (VQE)}
In this section, we extend the proposed 
multi-target compilation approach
into VQE. 
It is a quantum algorithm used to estimate the lowest eigenvalue of a given Hamiltonian, which is crucial in quantum chemistry and optimization problems. It employs a parameterized quantum circuit to prepare trial states and a classical optimization routine to adjust these parameters iteratively, converging towards the lowest eigenvalue. This algorithm is handy in problems where the direct calculation of eigenvalues is computationally expensive, making it a promising candidate for near-term quantum computer applications.

The unitary coupled cluster with singles and doubles (UCCSD) ansatz was used in the first proposal of the VQE algorithm \cite{Peruzzo2014}. While achieving high accuracy and attracting significant research interest, the UCCSD ansatz requires a large number of gates and a high depth \cite{Romero_2019}, due to their ``staircase'' structure. This makes it less suitable for the Noisy Intermediate-Scale Quantum (NISQ) era. 


\begin{figure}[t]
\centering
\includegraphics[width=8.6cm]{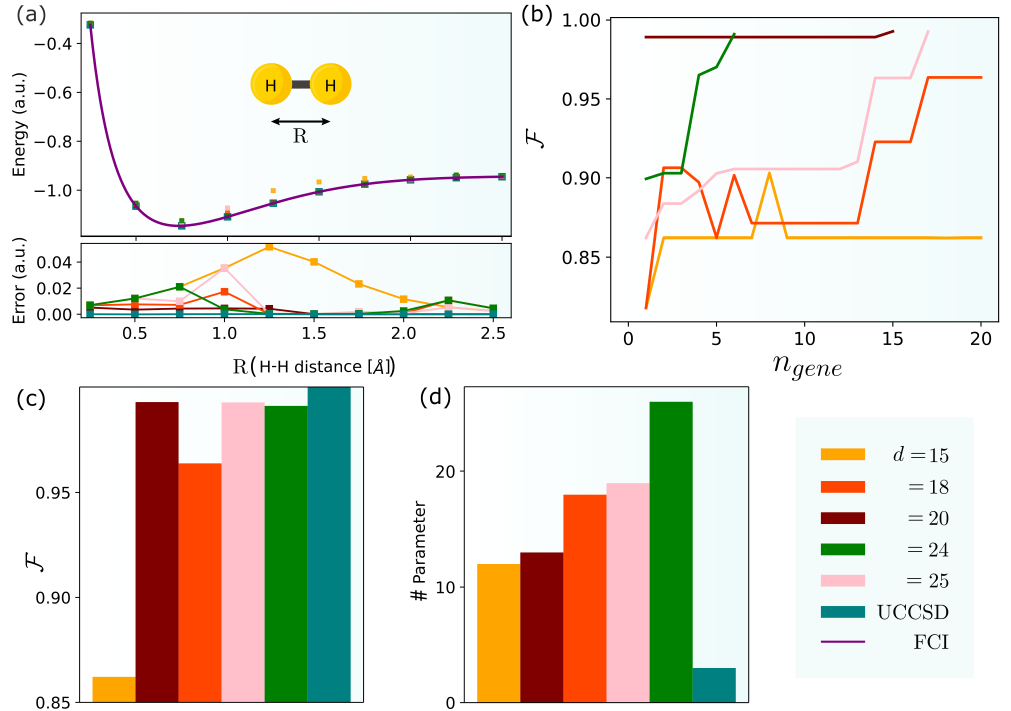}
\caption{
\textbf{The application of GA-VQA scheme to Variational Quantum Eigensolver Algorithm for Hydrogen molecule}
(a) Comparison of UCCSD ansatz and GA-VQA found ansatz in estimating ground state energy surface of the Hydrogen molecule.
(b) The fidelity for different depth GA-VQA found ansatz as a function of optimization step $n_{\rm gene} $.
(c) The best fidelity of GA-VQA found and UCCSD ansatz.
(d) The number of parameters in GA-VQA found and UCCSD ansatz.}
\label{fig:5}
\end{figure}

Here, we apply the developed GA-VQA to find a more efficient ansatz for electronic structure ansatz.
For this purpose, we use the cost function as the energy 
\begin{align}\label{eq:fi_vqe}
    \mathcal{L}_{\rm VQE}(\bm\Theta, V_l) 
    =\sum_{j=1}^{n}
    \langle\textbf{0}|V_l^\dagger(\bm\theta_j) H_jV_l(\bm\theta_j) 
    |\textbf{0}\rangle \;; \forall 1 \le l \le n_\mathbf{\mathcal{V}},
\end{align}
where $|\textbf{0}\rangle \equiv | 0 \rangle ^ {\otimes N}$ is a reference state, 
$H_j \in \mathcal{H}\ \forall j = \{1,\cdots,n\}$ is a set of $n$ target Hamiltonians
with respect to the molecular distances $R=\{R_1, R_2,\cdots, R_n\}$, 
and $V_l$ is a quantum circuit found by the GA-VQA scheme.

We apply the GA-VQA scheme to the Hydrogen molecule. In Figure ~\ref{fig:5}(a), we estimate the ground state potential energy surface by optimizing VQE with the best ansatz from the GA-VQA method for ten points ($n = 10$), achieving accuracy comparable to the UCCSD ansatz. Here, we approximate the UCCSD ansatz with only the first-order Suzuki-Trotter approximation. Figure \ref{fig:5}(b) shows the convergence of fidelity versus $n_{\rm gene} $ for $d =$ 15, 18, 20, 24, and 25, with their best fidelities plotted in Figure ~\ref{fig:5}(c). As seen in Figure ~\ref{fig:5}(d), the ansatzes identified by GA-VQA possess more parameters compared to UCCSD ansatz but require significantly less depth. Even with the first-order Trotter decomposition, the UCCSD ansatz demands a depth of 74. Consequently, the GA-VQA found ansatzes are more suitable for the NISQ era.
Details on the technical methods used in these findings are provided in \ref{appB}.

\section{Conclusion}\label{sec4}
We introduced a novel multi-target quantum compilation algorithm aimed at optimizing quantum circuits for multiple objectives simultaneously. By leveraging a genetic algorithm (GA) combined with variational quantum algorithms (VQAs), we were able to efficiently optimize both the structure and parameters of quantum circuits. Our benchmarks and case studies demonstrated the algorithm's capability to outperform traditional approaches, particularly in scenarios requiring simultaneous optimization, such as simulating thermal states and time-dependent systems. This method offers a significant advancement in quantum compilation, providing a foundation for further exploration in multi-target optimization and its potential applications in quantum computing.

Although this work focuses on developing the multi-target quantum compilation algorithm, analyzing its performance in noisy environments, especially on NISQ devices, is crucial. Noise can affect both the convergence of the genetic algorithm and the fidelity of the compiled circuits. In future works, we will include noise models to assess the algorithm's robustness and explore strategies to mitigate noise, enhancing convergence and maintaining high fidelity.

\begin{acknowledgments}
L.B.H. thanks Dr. Sahel Ashhab for the fruitful discussions. V.T.H. expresses gratitude to Dr. Luong Ngoc Hoang for advice on genetic algorithms. This research is funded by JSPS KAKENHI Grant Number 23K13025, Unitary Fund, and Tohoku University FRIS URO. L.N.T is partially supported by Vietnam National University Ho Chi Minh City (VNU-HCM) under grant number C2024-28-04.
\end{acknowledgments}

\section*{Data availability}
Data are available from the corresponding authors upon reasonable request.

\section*{Code availability}
The code is available at \url{https://github.com/vutuanhai237/GA-QAS}.

\section*{Author contributions statement}

V.T.H. wrote the GA-QAS code and conducted the numerical simulation. 
N.T.V. and J.U. implemented the thermal state preparation. 
J.U. and L.B.H. carried out the time-dependent dynamics simulation. 
N.V.L. implemented the VQE demo. 
L.N.T proposed and supervised the VQE demo. 
L.B.H. proposed the theory of multi-target quantum compilation and supervised its application to thermal state preparation and time-dependent dynamics simulation. 
L.N.T and L.B.H. supervised the entire project. 
All authors discussed the results and contributed to writing the manuscript.

\section*{Competing interests}
The author declares no competing interests.


\appendix

\setcounter{equation}{0}
\renewcommand{\theequation}{A.\arabic{equation}}
\section{Structure of GA}
\label{appA}
\subsection{GA scheme}
First, let us define some terminologies used in common genetic science and
their counterpart in quantum circuits. These terminologies are given in Tab.~\ref{tab:2} below.

\begin{table*}[ht]
\centering
\caption{List of terminologies used in the common genetic science and
their counterpart in quantum circuits.}
\begin{tabular}{lllll}\toprule
\hline
& \multicolumn{2}{c}{Genetic science} & \multicolumn{2}{c}{Quantum circuit}
\\\cmidrule(lr){2-3}\cmidrule(lr){4-5}
\hline
  & Name  & Description   & Name  & Description \\ \midrule
1 & DNA/RNA & genetic material & quantum gate  & unitary operator  \\
2 & individual & genetic unit & quantum circuit  & a set of quantum gates \\
3 & population & a group of genes & a set of quantum circuits  & a set of quantum circuits \\
4 & fitness & performance metric & cost function  &  performance metric  \\ \bottomrule
\end{tabular}
\label{tab:2}
\end{table*}
In a GA scheme, each quantum circuit (individual gene) 
will be randomly generated from a pool gate 
with a fixed circuit depth, 
which consists of various types of 
quantum gates with one qubit, two qubits, to one-parameter, and two-parameter. 
The more gates used, the more possible candidates. 
This increases the likelihood of finding a suitable 
candidate but at the same time expands the search space, 
making us spend more time. Furthermore, 
using basic gates such as 
$\text{Clifford set} = \{{\rm H, S, CX}\}$ \cite{gottesman1998theory} 
enables to implementation of candidates 
in a real quantum computer, two-qubit gates 
will be used restrictively. In general, our pool is $\{{\rm H,S, CX,R}_i(.)\}$ with $i\in\{x,y,z\}$. 
Although the H and S gates can be expressed using \( R_i(.) \), they are essential for reducing the complexity of the ansatz, including lowering the number of parameters and minimizing circuit depth.

The scheme will produce a set of quantum circuits and evaluate their fitness using a fitness function to identify the optimal quantum circuit. If the best fitness falls short of a predetermined threshold, an evolutionary process involving selection, crossover, and mutation will be implemented to generate a new circuit. This process will be repeated iteratively until the threshold is met or till the end number of generation, we then switch to VQA to optimize parameters $\bm\Theta$.

\subsection{Selection-Crossover-Mutation}
There are many types of selection, crossover, and mutation functions:
\begin{itemize}
    \item Selection: Tournament, Proportional, Rank, Elitist,...
    \item Crossover: One-point, N-point, Uniform, Linear combination,...
    \item Mutation: Random deviation, Exchange, Shift, Bit flip, Inversion, Shuffle,...
\end{itemize}
In this work, we used Elitist Selection, 
One-point Crossover, and Bit Flip Mutation 
as the default option. We plan to investigate 
other combinations in future studies.
After generating a set of quantum circuits $\mathcal{V}$, 
we evaluate them using cost functions like fidelity 
and retain only the two best candidates as 
elitist circuits for the next generation. 
These selected candidates are paired for crossover to create new ones.
Each pair of parents ($p_1,p_2$) is divided into four parts 
$\{p_{11}, p_{12}, p_{21}, p_{22}\}$ at one point, 
normally center point. Two new candidates ($c_1, c_2$) 
are formed by combining these parts: $c_1=\{p_{11}, 
p_{22}\}$ and $c_2=\{p_{21}, p_{12}\}$. In each generation, 
there is a small probability ($1\%$) that any gate (bit) 
in a candidate will mutate and be replaced (flip) 
with a different gate from the pool.

\setcounter{equation}{0}
\renewcommand{\theequation}{B.\arabic{equation}}
\section{Detailed experimental setting}
\label{appB}
All numerical findings are implemented in Python using Qiskit 0.45.1 with QASMSimulatorPy simulation to verify algorithm convergence. We benchmark $2$ to $10$ qubits on several computer systems, including 
24 nodes CPU Intel Xeon X5675 High-performance computer (HPC) at Vietnam Academy Science \& Technology, Intel X299-GPU A6000 Workstation and AMD EPYC 7713P cluster at Tohoku University, and Intel Core i9-10940X CPU at NAIST, Japan.

We mainly use gradient-based method with Adam optimizer (except for the VQE), 
where a set of $\bm \theta$ is updated through
\begin{align}\label{eq:adam}
\bm{\theta}^{k+1}=\bm{\theta}^{k}
-\alpha\frac{\hat{m}_{k}}{\sqrt{\hat{v}_{k}} + \epsilon},
\end{align}
where $m_{k}=\beta_{1} m_{k-1}
+\left(1-\beta_{1}\right) 
\nabla_{\bm\theta}\mathcal{L}(\bm\theta), 
v_{k}=\beta_{2} v_{k-1}+(1-\beta_{2}) 
\nabla_{\bm\theta}^2\mathcal{L}(\bm\theta),
\hat{m}_{k}=m_{k} /\left(1-\beta_{1}^{k}\right),
\hat{v}_{k}=v_{k} /\left(1-\beta_{2}^{k}\right),
$
with the hyper-parameters 
are chosen as 
$\alpha = 0.2, \beta_1 = 0.8, 
\beta_2 = 0.999$ 
and $\epsilon = 10^{-8}$. 
The gradient $\partial_{\bm\theta}\mathcal{L}(\bm\theta)$
is given through the general parameter-shift rule
\cite{Hai2024}.

\subsection{Numerical benchmarking}

For numerical benchmarking, we generated Haar random unitaries, using 20 for training and 10 for testing at each time. We ran GA-VQA on the training set with:
\begin{itemize}
    \item $n_{\rm iter}  = 100$,
    \item threshold = 0.01,
    \item $n_{\rm gene} $ ranging from 10 to 40,
    \item $n_\mathcal{V} = [4, 8, 16, 32]$,
    \item Depth $d$ from 2 to 39,
\end{itemize}
and get the best circuits for each setup. We use these best circuits to calculate the risk on the testing set. All results in Figure ~\ref{fig:2} are for the testing set.

\subsection{Thermal state preparation}
\subsubsection{Conventional purification method.}
For the conventional purification method, 
we examine for $N = 2$ thermal state, 
which requires a 4-qubit circuit. Initially, 
we generate 100 target TFD states corresponding 
to equally-spaced $\beta \in [0,10]$. 
We then run GA-VQA to find the best circuit 
structure for all those TFD states based on 
the weighted-sum cost function as we will 
explain below. Finally, we continue to run 
optimization using Adam optimizer for 100 iterations to find optimal parameters for each quantum state.

The circuit depth is 29 after transpiling. 
Other configurations include $n_{\rm iter}  
= 100$, $n_\mathcal{V} = 16$, $n_{\rm gene}  = 20$.
Furthermore, optimal parameters with $\beta < 1$ 
are harder to find than those with $\beta > 1$. 
A simple average of cost values in the interval 
$\beta$ from 0 to 10 is not sufficient for 
finding near-optimal ansatzes in the range  
$\beta$ from 0 to 1. To address this, 
we use a weighted average cost function 
for different intervals. Specifically, 
we assign weights $w_1=2.2,w_2=1.6,w_3=0.9$ 
to three corresponding intervals 
$\beta \in [0,4),\beta \in [4,7),$ 
and $\beta \in [7,10]$. The weighted-sum 
cost function is expressed as
\begin{align}\label{eq:FitnessFunction}
    \mathcal{F}_{\rm ws} = 
    \frac{w_1*\mathcal{F}_{\beta \in [0,4)}+w_2*
    \mathcal{F}_{\beta \in [4,7)}+w_3*
    \mathcal{F}_{\beta \in [7,10]}}{w_1+w_2+w_3},
\end{align}
where $\mathcal{F}_{\beta \in [i,j)}, 
0 \leq i ,j \leq 10 $ is the average 
fidelity in the interval $[i,j)$.

\subsubsection{Dense-purification method.}
We examine thermal state preparation using the dense-purification method across 2 to 8 qubits. Parameters include $n_{\rm iter}  = 100$, $d = 2N$, 
$n_\mathcal{V} = 8$, $n_{\rm gene}  = 16$. 
Note that $n_{\rm gene} $ does not significantly affect the runtime, as the program can stop anytime if the threshold is met.
The threshold and optimizer remain consistent with the previous settings.
Note that in this case, the cost function is 
fidelity $\mathcal{F}$ in Eq.~\eqref{eq:GibbsFidelity}.

\begin{table*}[ht]
\centering
\caption{Experiment setting}
\begin{tabular}{lllll}\toprule
\hline
Application & Benchmarking & TSP & TD-QDS & VQE\\ \hline
Cost function & $\mathcal{L} \eqref{eq:f_qc}$ & $\mathcal{F} \eqref{eq:GibbsFidelity}, \mathcal{F}_{\text{ws}} \eqref{eq:FitnessFunction}$ &  
 $\mathcal{L}^2 \eqref{eq:method:L2}$ & $\mathcal{L}_{\rm VQE} \eqref{eq:fi_vqe}$\\ 
Evaluated object & Haar random unitary & Thermal state & Heisenberg models & $\text{H}_2$ molecules\\
Optimizer & Adam & Adam & Adam & COBYLA\\
$N$ & $[2-5]$ & [2-8] & $2$ & 4\\
\hline
\multirow{3}{*}{Hyper-parameter} & $d=[2-39]$  & $d = 2N$ & $d=4$ & d=[15,18,20,24,25]\\
                                & $n_\mathcal{V}=[4,8,16,32]$  & $n_\mathcal{V}=8$  & $n_\mathcal{V}=8$ & $n_\mathcal{V}=8$ \\
                                & $n_{\rm gene} =[10, 20, 30, 40]$  & $n_{\rm gene}  = 16$
                                &  $n_{\rm gene} =16$   
                                &  $n_{\rm gene} =20$ \\
\bottomrule
\end{tabular}%
\label{tab:experiment}
\end{table*}

\subsection{Time-dependent dynamic simulation}
\subsubsection{GA-VQA method.}
For the GA-VQA method, our simulation follows these steps:
(1) Create the initial state $|\psi_0\rangle = |11...00\rangle$
by applying X gates to $N/2$ qubits.
(2) Create a set of 100 target unitaries $U(t_i)$ using Eq.~\eqref{eq:UXYZ} for $t_i \in [0, 10],\ i = \{1, 2, \cdots, 100\}$.
(3) We then run GA-VQA to find the best circuit structure for all those target unitaries.
Once $V^*(\bm\Theta^*)$ is determined, 
we measure the magnetization as follows.
(i) Prepare $|\psi_0\rangle$ in a quantum circuit.
(ii) Apply $V^*(\bm\Theta^*)$ to get the final state
$|\check\psi(t)\rangle = V^* (\bm\Theta^*)|\psi_0\rangle$.
(iii) Measure qubit $j$ and get the probability
\begin{align}
    p_j(m) = \langle\check\psi(t)|\Pi_j|\check\psi(t)\rangle,
    \text{ for } m\in\{0, 1\},
\end{align}
where $\Pi_j = I_1\otimes\cdots\otimes
|m_j\rangle\langle m_j|\otimes\cdots\otimes I_N$.
Finally, the local magnetization \eqref{eq:M_theo} is given by
\begin{align}\label{eq:M_ga}
M_j(t) =  p_j(0) - p_j(1).
\end{align}
We benchmark for 2-qubit case. The parameters used for this case include: $n_{\rm iter}  = 500$, $d = 4$, 
$n_\mathcal{V} = 8$, $n_{\rm gene}  = 16$,
and the local magnetization was taken for the second qubit. 
See Tab.~\ref{tab:experiment} for a summary.

\subsubsection{Trotterization method.}
To implement the Trotterization, we break down the evolution $U(t_i)$ as follows
\begin{widetext}
\begin{align}\label{eq:method:Ut}
\notag U(t_{i}) &= \exp\Big[-i\int_{t_{i-1}}^{t_i}H(s)ds\Big]\\
 &= \exp\Bigg\{-i\frac{\delta t}{K}\lim_{K\to\infty}
    \Big[H\big(t_{i-1}\big) + H\big(t_{i-1}+\frac{\delta t}{K}\big) + \cdots + H\big(t_{i-1}+\frac{(K-1)\delta t}{K}\big)\Big]\Bigg\},
\end{align}
\end{widetext}
where $\delta t = t_i - t_{i-1} = 0.1$ is the interval, divided into $K$ equal sub-intervals. Here, we set $K = 5$, making $\delta t/K = 0.02$ sufficiently small. Each $H(s)$ in the above expression consists of single and two-qubit gates, and we assume $H(s) = \sum_{j=1}^LH_j$.
Using Trotterization, we can break $H(s)$
own into a sequence of quantum gates and implement it in quantum circuits. The Trotterization expansions of the first order (Trotter1) and second order (Trotter2) are given by
\cite{PhysRevLett.128.210501,ikeda2023trotter24}
\begin{widetext}
\begin{align}\label{qe:appTrot}
    e^{-iH(s)} &= \lim_{r\to\infty} \Big(e^{-i\frac{H_L}{r}}\cdots e^{-i\frac{H_1}{r}}\Big)^r \text{ for Trotter1, and } \\
    e^{-iH(s)} &= \lim_{r\to\infty} \Big[\Big(e^{-i\frac{H_L}{2r}}\cdots e^{-i\frac{H_1}{2r}}\Big)
    \Big(e^{-i\frac{H_1}{2r}}\cdots e^{-i\frac{H_L}{2r}}\Big)\Big]^r \text{ for Trotter2},
\end{align}
\end{widetext}
where $r$ is the Trotter number. 
In this work, we set $r = 100$.

\subsubsection{AVQDS method.}
The AVQDS method, introduced in 
Ref.~\cite{PRXQuantum.2.030307}, 
works as follows.
Starting with the initial quantum state $|\psi_0\rangle$, 
the state evolves over time according to
$|\psi(t)\rangle$ as
\begin{align}\label{eq:method:schro}
    \dfrac{d|\psi(t)\rangle}{dt} = -i H(t)|\psi(t)\rangle.
\end{align}
To solve this, we use a variational quantum ansatz $|\psi(\bm\theta(t)\rangle$,
where $\bm\theta(t)$ represents a set of time-dependent parameters. For example, at time $t = 0$, it gives
$|\psi(\bm\theta(t = 0))\rangle = |\psi_0\rangle$.
These parameters are trained 
to minimize the squared McLachlan distance \cite{doi:10.1080/00268976400100041} $\mathcal{L}^2$
\begin{align}\label{eq:method:L2}
    \mathcal{L}^2 = \Big\|
    \sum_\mu\dfrac{d|\psi(\bm\theta)\rangle}{d\theta_\mu}
    \dfrac{d\theta_\mu}{dt} +i H(t)|\psi(\bm\theta)\rangle\Big\|^2_F,
\end{align}
where $\|\cdot\|_F$ denotes the Frobenius norm, 
and we omit $t$ in $\bm\theta(t)$ to simplify the notation.

In our numerical simulation, we first construct the Hamiltonian
$H(t)$ as described in Eq.~\eqref{eq:XYZ_model}. 
We then employ the pseudo-Trotter ansatz \cite{PRXQuantum.2.030307} 
$|\psi(\bm\theta)\rangle 
= \prod_{\mu = 1}^{N_{\bm\theta}}e^{-i\theta_\mu A_\mu}|\psi_0\rangle$,
where $N_{\bm\theta}$ is the number of trainable time-dependent parameters, and $A_\mu$ are Hermitian operators from a set of Pauli operators. 
During the simulation for any time $0 < t_i < T$, the ansatz is adaptively updated until the squared McLachlan distance meets the threshold $\mathcal{L}^2_{\rm cut}$.
We set the threshold $\mathcal{L}^2_{\rm cut} = 10^{-3}$ for Figure ~\ref{fig:4}(a), 
and $\mathcal{L}^2_{\rm cut} = 10^{-1}$ for Figure ~\ref{fig:4}(b,c).
In Figure ~\ref{fig:4}(b,c), 
the AVQDS method is not trainable at $\mathcal{L}^2_{\rm cut} = 10^{-3}$,
so we temporarily reduce its accuracy by training at 
$\mathcal{L}^2_{\rm cut} = 10^{-1}$. This is why the accuracy of this method is lower compared to the exact result and GA-VQA method.

\subsection{Variational quantum eigensolver}
For the variational quantum eigensolver \cite{Peruzzo2014,TILLY20221}, 
an upper bound of ground state energy $E_o$ of a given Hamiltonian is bounded by
\begin{equation}
    E_o \leq \frac{\langle \psi ( \theta) | H | \psi (\theta) \rangle}{\langle \psi (\theta) | \psi ( \theta )\rangle} \equiv \langle H \rangle,
\end{equation}
which is found by optimizing the parameters of quantum state $\displaystyle |\psi \rangle \equiv V_l (\theta) |\textbf{0}\rangle$. To simulate the GA-VQA scheme's application in VQE with a Hydrogen molecule, we run GA-VQA with the setup that can be seen in the VQE field at Tab. \ref{tab:experiment}. In this case, the target of the GA-VQA is to find ansatzes with low depth but still gain the needed accuracy of ground state molecular energy. The GA is leveraged to find the friendly NISQ ansatz while the optimizing part is left to VQE. This method iteratively converges to higher fidelity (the higher the fidelity, the more accurate the result) until the max generation or termination condition is satisfied.

\bibliographystyle{unsrt}
\bibliography{refs}

\begin{thebibliography}{10}

\bibitem{Cerezo2021}
M.~Cerezo, Andrew Arrasmith, Ryan Babbush, Simon~C. Benjamin, Suguru Endo,
  Keisuke Fujii, Jarrod~R. McClean, Kosuke Mitarai, Xiao Yuan, Lukasz Cincio,
  and Patrick~J. Coles.
\newblock Variational quantum algorithms.
\newblock {\em Nature Reviews Physics}, 3(9):625--644, Sep 2021.

\bibitem{Kuzmin2020variationalquantum}
Viacheslav~V. Kuzmin and Pietro Silvi.
\newblock Variational quantum state preparation via quantum data buses.
\newblock {\em {Quantum}}, 4:290, July 2020.

\bibitem{Sagastizabal2021}
R.~Sagastizabal, S.~P. Premaratne, B.~A. Klaver, M.~A. Rol, V.~Neg{\^i}rneac,
  M.~S. Moreira, X.~Zou, S.~Johri, N.~Muthusubramanian, M.~Beekman,
  C.~Zachariadis, V.~P. Ostroukh, N.~Haider, A.~Bruno, A.~Y. Matsuura, and
  L.~DiCarlo.
\newblock Variational preparation of finite-temperature states on a quantum
  computer.
\newblock {\em npj Quantum Information}, 7(1):130, Aug 2021.

\bibitem{PhysRevLett.129.230504}
Xiao-Ming Zhang, Tongyang Li, and Xiao Yuan.
\newblock Quantum state preparation with optimal circuit depth: Implementations
  and applications.
\newblock {\em Phys. Rev. Lett.}, 129:230504, Nov 2022.

\bibitem{castro2024variational}
Juan C.~Zuñiga Castro, Jeffrey Larson, Sri Hari~Krishna Narayanan, Victor~E.
  Colussi, Michael~A. Perlin, and Robert~J. Lewis-Swan.
\newblock Variational quantum state preparation for quantum-enhanced metrology
  in noisy systems, 2024.

\bibitem{hai2023variational}
Vu~Tuan Hai, Nguyen~Tan Viet, and Le~Bin Ho.
\newblock Variational preparation of entangled states on quantum computers.
\newblock {\em arXiv preprint arXiv:2306.17422}, 2023.

\bibitem{HAI2024101726}
Vu~Tuan Hai, Nguyen~Tan Viet, and Le~Bin Ho.
\newblock <qo|op>: A quantum object optimizer.
\newblock {\em SoftwareX}, 26:101726, 2024.

\bibitem{PRXQuantum.2.030307}
Yong-Xin Yao, Niladri Gomes, Feng Zhang, Cai-Zhuang Wang, Kai-Ming Ho, Thomas
  Iadecola, and Peter~P. Orth.
\newblock Adaptive variational quantum dynamics simulations.
\newblock {\em PRX Quantum}, 2:030307, Jul 2021.

\bibitem{Luo2024}
Jianming Luo, Kaihan Lin, and Xing Gao.
\newblock Variational quantum simulation of lindblad dynamics via quantum state
  diffusion.
\newblock {\em The Journal of Physical Chemistry Letters}, 15(13):3516--3522,
  Apr 2024.

\bibitem{PhysRevResearch.6.023130}
David Linteau, Stefano Barison, Netanel~H. Lindner, and Giuseppe Carleo.
\newblock Adaptive projected variational quantum dynamics.
\newblock {\em Phys. Rev. Res.}, 6:023130, May 2024.

\bibitem{Koczor_2020}
Bálint Koczor, Suguru Endo, Tyson Jones, Yuichiro Matsuzaki, and Simon~C
  Benjamin.
\newblock Variational-state quantum metrology.
\newblock {\em New Journal of Physics}, 22(8):083038, aug 2020.

\bibitem{9605341}
Ziqi Ma, Pranav Gokhale, Tian-Xing Zheng, Sisi Zhou, Xiaofei Yu, Liang Jiang,
  Peter Maurer, and Frederic~T. Chong.
\newblock Adaptive circuit learning for quantum metrology.
\newblock In {\em 2021 IEEE International Conference on Quantum Computing and
  Engineering (QCE)}, pages 419--430, 2021.

\bibitem{Meyer2021}
Johannes~Jakob Meyer, Johannes Borregaard, and Jens Eisert.
\newblock A variational toolbox for quantum multi-parameter estimation.
\newblock {\em npj Quantum Information}, 7(1):89, Jun 2021.

\bibitem{Le2023}
Trung~Kien Le, Hung~Q. Nguyen, and Le~Bin Ho.
\newblock Variational quantum metrology for multiparameter estimation under
  dephasing noise.
\newblock {\em Scientific Reports}, 13(1):17775, Oct 2023.

\bibitem{Cimini2024}
Valeria Cimini, Mauro Valeri, Simone Piacentini, Francesco Ceccarelli, Giacomo
  Corrielli, Roberto Osellame, Nicol{\`o} Spagnolo, and Fabio Sciarrino.
\newblock Variational quantum algorithm for experimental photonic
  multiparameter estimation.
\newblock {\em npj Quantum Information}, 10(1):26, Feb 2024.

\bibitem{PRXQuantum.5.020332}
Zhiyan Ding, Xiantao Li, and Lin Lin.
\newblock Simulating open quantum systems using hamiltonian simulations.
\newblock {\em PRX Quantum}, 5:020332, May 2024.

\bibitem{heya2018variational}
Kentaro Heya, Yasunari Suzuki, Yasunobu Nakamura, and Keisuke Fujii.
\newblock Variational quantum gate optimization, 2018.

\bibitem{Khatri2019quantumassisted}
Sumeet Khatri, Ryan LaRose, Alexander Poremba, Lukasz Cincio, Andrew~T.
  Sornborger, and Patrick~J. Coles.
\newblock Quantum-assisted quantum compiling.
\newblock {\em {Quantum}}, 3:140, May 2019.

\bibitem{PRXQuantum.2.040327}
Tyler Volkoff, Zo\'e Holmes, and Andrew Sornborger.
\newblock Universal compiling and (no-)free-lunch theorems for
  continuous-variable quantum learning.
\newblock {\em PRX Quantum}, 2:040327, Nov 2021.

\bibitem{hai2023universal}
Vu~Tuan Hai and Le~Bin Ho.
\newblock Universal compilation for quantum state tomography.
\newblock {\em Scientific Reports}, 13(1):3750, Mar 2023.

\bibitem{PhysRevA.99.042314}
M\'aria Kieferov\'a, Artur Scherer, and Dominic~W. Berry.
\newblock Simulating the dynamics of time-dependent hamiltonians with a
  truncated dyson series.
\newblock {\em Phys. Rev. A}, 99:042314, Apr 2019.

\bibitem{Berry2024quantumalgorithm}
Dominic~W. Berry and Pedro C.~S.~Costa.
\newblock Quantum algorithm for time-dependent differential equations using
  {D}yson series.
\newblock {\em {Quantum}}, 8:1369, June 2024.

\bibitem{Mizuta2023optimalhamiltonian}
Kaoru Mizuta and Keisuke Fujii.
\newblock Optimal {H}amiltonian simulation for time-periodic systems.
\newblock {\em {Quantum}}, 7:962, March 2023.

\bibitem{PhysRevLett.123.220502}
Jingxiang Wu and Timothy~H. Hsieh.
\newblock Variational thermal quantum simulation via thermofield double states.
\newblock {\em Phys. Rev. Lett.}, 123:220502, Nov 2019.

\bibitem{Zhu_2020}
D.~Zhu, S.~Johri, N.~M. Linke, K.~A. Landsman, C.~Huerta~Alderete, N.~H.
  Nguyen, A.~Y. Matsuura, T.~H. Hsieh, and C.~Monroe.
\newblock Generation of thermofield double states and critical ground states
  with a quantum computer.
\newblock {\em Proceedings of the National Academy of Sciences},
  117(41):25402–25406, September 2020.

\bibitem{PhysRevApplied.16.054035}
Youle Wang, Guangxi Li, and Xin Wang.
\newblock Variational quantum gibbs state preparation with a truncated taylor
  series.
\newblock {\em Phys. Rev. Appl.}, 16:054035, Nov 2021.

\bibitem{PhysRevA.90.053818}
Brandon~S. Harms, Blake~E. Anthony, Noah~T. Holte, Hunter~A. Dassonville, and
  Andrew M.~C. Dawes.
\newblock Multimode quantum state tomography using unbalanced array detection.
\newblock {\em Phys. Rev. A}, 90:053818, Nov 2014.

\bibitem{He2024}
Kevin He, Ming Yuan, Yat Wong, Srivatsan Chakram, Alireza Seif, Liang Jiang,
  and David~I. Schuster.
\newblock Efficient multimode wigner tomography.
\newblock {\em Nature Communications}, 15(1):4138, May 2024.

\bibitem{Ikeda2023minimum}
Tatsuhiko~N. Ikeda, Asir Abrar, Isaac~L. Chuang, and Sho Sugiura.
\newblock Minimum {T}rotterization {F}ormulas for a {T}ime-{D}ependent
  {H}amiltonian.
\newblock {\em {Quantum}}, 7:1168, November 2023.

\bibitem{ikeda2023trotter24}
Tatsuhiko~N. Ikeda, Hideki Kono, and Keisuke Fujii.
\newblock Trotter24: A precision-guaranteed adaptive stepsize trotterization
  for hamiltonian simulations, 2023.

\bibitem{Du2022}
Yuxuan Du, Tao Huang, Shan You, Min-Hsiu Hsieh, and Dacheng Tao.
\newblock Quantum circuit architecture search for variational quantum
  algorithms.
\newblock {\em npj Quantum Information}, 8(1):62, May 2022.

\bibitem{Zhang_2021}
Shi-Xin Zhang, Chang-Yu Hsieh, Shengyu Zhang, and Hong Yao.
\newblock Neural predictor based quantum architecture search.
\newblock {\em Machine Learning: Science and Technology}, 2(4):045027, oct
  2021.

\bibitem{PhysRevResearch.2.023074}
Li~Li, Minjie Fan, Marc Coram, Patrick Riley, and Stefan Leichenauer.
\newblock Quantum optimization with a novel gibbs objective function and ansatz
  architecture search.
\newblock {\em Phys. Rev. Res.}, 2:023074, Apr 2020.

\bibitem{https://doi.org/10.1002/qute.202100134}
Zhimin He, Chuangtao Chen, Lvzhou Li, Shenggen Zheng, and Haozhen Situ.
\newblock Quantum architecture search with meta-learning.
\newblock {\em Advanced Quantum Technologies}, 5(8):2100134, 2022.

\bibitem{Grimsley2019}
Harper~R. Grimsley, Sophia~E. Economou, Edwin Barnes, and Nicholas~J. Mayhall.
\newblock An adaptive variational algorithm for exact molecular simulations on
  a quantum computer.
\newblock {\em Nature Communications}, 10(1):3007, Jul 2019.

\bibitem{934383}
B.I.P. Rubinstein.
\newblock Evolving quantum circuits using genetic programming.
\newblock In {\em Proceedings of the 2001 Congress on Evolutionary Computation
  (IEEE Cat. No.01TH8546)}, volume~1, pages 144--151 vol. 1, 2001.

\bibitem{8862255}
Annu Lambora, Kunal Gupta, and Kriti Chopra.
\newblock Genetic algorithm- a literature review.
\newblock In {\em 2019 International Conference on Machine Learning, Big Data,
  Cloud and Parallel Computing (COMITCon)}, pages 380--384, 2019.

\bibitem{katoch2021review}
Sourabh Katoch, Sumit~Singh Chauhan, and Vijay Kumar.
\newblock A review on genetic algorithm: past, present, and future.
\newblock {\em Multimedia Tools and Applications}, 80(5):8091--8126, Feb 2021.

\bibitem{Tandeitnik_2024}
Daniel Tandeitnik and Thiago Guerreiro.
\newblock Evolving quantum circuits.
\newblock {\em Quantum Information Processing}, 23(3), March 2024.

\bibitem{Ostaszewski2021structure}
Mateusz Ostaszewski, Edward Grant, and Marcello Benedetti.
\newblock Structure optimization for parameterized quantum circuits.
\newblock {\em {Quantum}}, 5:391, January 2021.

\bibitem{10.1145/3550488}
Ang Li and et~al.
\newblock Qasmbench: A low-level quantum benchmark suite for nisq evaluation
  and simulation.
\newblock {\em ACM Transactions on Quantum Computing}, 4(2), feb 2023.

\bibitem{Creevey2023}
Floyd~M. Creevey, Charles~D. Hill, and Lloyd C.~L. Hollenberg.
\newblock Gasp: a genetic algorithm for state preparation on quantum
  computers.
\newblock {\em Scientific Reports}, 13(1):11956, Jul 2023.

\bibitem{PhysRevA.106.022426}
Sahel Ashhab, Naoki Yamamoto, Fumiki Yoshihara, and Kouichi Semba.
\newblock Numerical analysis of quantum circuits for state preparation and
  unitary operator synthesis.
\newblock {\em Phys. Rev. A}, 106:022426, Aug 2022.

\bibitem{PhysRevA.109.052605}
Sahel Ashhab, Fumiki Yoshihara, Miwako Tsuji, Mitsuhisa Sato, and Kouichi
  Semba.
\newblock Quantum circuit synthesis via a random combinatorial search.
\newblock {\em Phys. Rev. A}, 109:052605, May 2024.

\bibitem{Caro2023}
Matthias~C. Caro, Hsin-Yuan Huang, Nicholas Ezzell, Joe Gibbs, Andrew~T.
  Sornborger, Lukasz Cincio, Patrick~J. Coles, and Zo{\"e} Holmes.
\newblock Out-of-distribution generalization for learning quantum dynamics.
\newblock {\em Nature Communications}, 14(1):3751, Jul 2023.

\bibitem{ezzell2023quantum}
Nic Ezzell, Elliott~M Ball, Aliza~U Siddiqui, Mark~M Wilde, Andrew~T
  Sornborger, Patrick~J Coles, and Zo{\'e} Holmes.
\newblock Quantum mixed state compiling.
\newblock {\em Quantum Science and Technology}, 8(3):035001, 2023.

\bibitem{9571985}
Jacob~M. Leamer, Wenlei Zhang, Ravi~K. Saripalli, Ryan~T. Glasser, and Denys~I.
  Bondar.
\newblock Simulation of quantum gibbs states using epsilon-near-zero materials
  and classical light.
\newblock In {\em Conference on Lasers and Electro-Optics}, page JTu3A.72.
  Optica Publishing Group, 2021.

\bibitem{PhysRevX.8.021050}
Mohammad~H. Amin, Evgeny Andriyash, Jason Rolfe, Bohdan Kulchytskyy, and Roger
  Melko.
\newblock Quantum boltzmann machine.
\newblock {\em Phys. Rev. X}, 8:021050, May 2018.

\bibitem{PRXQuantum.4.010305}
Luuk Coopmans, Yuta Kikuchi, and Marcello Benedetti.
\newblock Predicting gibbs-state expectation values with pure thermal shadows.
\newblock {\em PRX Quantum}, 4:010305, Jan 2023.

\bibitem{Lewin2021}
Mathieu Lewin, Phan~Th{\`a}nh Nam, and Nicolas Rougerie.
\newblock Classical field theory limit of many-body quantum gibbs states in 2d
  and 3d.
\newblock {\em Inventiones mathematicae}, 224(2):315--444, May 2021.

\bibitem{ISRAEL1976107}
W.~Israel.
\newblock Thermo-field dynamics of black holes.
\newblock {\em Physics Letters A}, 57(2):107--110, 1976.

\bibitem{Gao2017}
Ping Gao, Daniel~Louis Jafferis, and Aron~C. Wall.
\newblock Traversable wormholes via a double trace deformation.
\newblock {\em Journal of High Energy Physics}, 2017(12):151, Dec 2017.

\bibitem{10.21468/SciPostPhys.6.3.029}
Wen~Wei Ho and Timothy~H. Hsieh.
\newblock {Efficient variational simulation of non-trivial quantum states}.
\newblock {\em SciPost Phys.}, 6:029, 2019.

\bibitem{Premaratne_2020}
Shavindra~P. Premaratne and A.~Y. Matsuura.
\newblock Engineering a cost function for real-world implementation of a
  variational quantum algorithm.
\newblock In {\em 2020 IEEE International Conference on Quantum Computing and
  Engineering (QCE)}. IEEE, 10 2020.

\bibitem{zhao2023adaptive}
Hongzheng Zhao, Marin Bukov, Markus Heyl, and Roderich Moessner.
\newblock Adaptive trotterization for time-dependent hamiltonian quantum
  dynamics using instantaneous conservation laws, 2023.

\bibitem{Peruzzo2014}
Alberto Peruzzo, Jarrod McClean, Peter Shadbolt, Man-Hong Yung, Xiao-Qi Zhou,
  Peter~J. Love, Al{\'a}n Aspuru-Guzik, and Jeremy~L. O'Brien.
\newblock A variational eigenvalue solver on a photonic quantum processor.
\newblock {\em Nature Communications}, 5(1):4213, Jul 2014.

\bibitem{Romero_2019}
Jonathan Romero, Ryan Babbush, Jarrod~R McClean, Cornelius Hempel, Peter~J
  Love, and Alán Aspuru-Guzik.
\newblock Strategies for quantum computing molecular energies using the unitary
  coupled cluster ansatz.
\newblock {\em Quantum Science and Technology}, 4(1):014008, oct 2018.

\bibitem{gottesman1998theory}
Daniel Gottesman.
\newblock Theory of fault-tolerant quantum computation.
\newblock {\em Phys. Rev. A}, 57:127--137, Jan 1998.

\bibitem{Hai2024}
Vu~Tuan Hai and Le~Bin Ho.
\newblock {\em Lagrange Interpolation Approach for General Parameter-Shift
  Rule}, pages 1--17.
\newblock Springer International Publishing, Cham, 2024.

\bibitem{PhysRevLett.128.210501}
David Layden.
\newblock First-order trotter error from a second-order perspective.
\newblock {\em Phys. Rev. Lett.}, 128:210501, May 2022.

\bibitem{doi:10.1080/00268976400100041}
A.D. McLachlan.
\newblock A variational solution of the time-dependent schrodinger equation.
\newblock {\em Molecular Physics}, 8(1):39--44, 1964.

\bibitem{TILLY20221}
Jules Tilly, Hongxiang Chen, Shuxiang Cao, Dario Picozzi, Kanav Setia, Ying Li,
  Edward Grant, Leonard Wossnig, Ivan Rungger, George~H. Booth, and Jonathan
  Tennyson.
\newblock The variational quantum eigensolver: A review of methods and best
  practices.
\newblock {\em Physics Reports}, 986:1--128, 2022.
\newblock The Variational Quantum Eigensolver: a review of methods and best
  practices.

\end{thebibliography}
\end{document}